\documentclass[a4paper]{jpconf}
\usepackage{graphicx,graphics}
\begin{document}
\title{Detection of discretized single-shell penetration in mesoscopic vortex matter}

\author{ M. I. Dolz $^1$,   Y. Fasano $^2$, N. R. Cejas Bolecek $^2$, H. Pastoriza $^2$, M. Konczykowski $^3$, and  C. J. van der Beek $^3$}
\address{$^1$ Departamento de F\'{i}sica, Universidad Nacional de San Luis, CONICET, Argentina}
\address{$^2$ Laboratorio de Bajas Temperaturas, Centro At\'{o}mico Bariloche, CNEA, Bariloche, Argentina}
\address{$^3$ Laboratoire des Solides Irradi\'{e}s, Ecole Polytechnique, CNRS URA-1380, Palaiseau, France}

\ead{$^1$ mdolz@unsl.edu.ar}

\begin{abstract}

We investigated  configurational changes in mesoscopic vortex
matter with less than thousand vortices during flux penetration in
freestanding
 50\,$\mu$m diameter
disks of Bi$_{2}$Sr$_{2}$CaCu$_{2}$O$_{8 +\delta}$.
High-resolution AC and DC local magnetometry data reveal
oscillations in the transmittivity echoed in peaks in the
third-harmonics magnetic signal fainting on increasing vortex
density. By means of extra experimental evidence and a simple
geometrical analysis we show that these features fingerprint the
discretized entrance of single-shells of vortices having a shape
that mimics the sample edge.
\end{abstract}

\section{Introduction}

The study of vortex configurations, flux penetration, multi-vortex
and giant-vortex states stabilized by confinement  have attracted
much attention during the last decade
\cite{Moshchalkov95,Geim97,Schweigert98b,Palacios98,Bruyndoncx99,Cabral04}.
The vortex arrangement in  superconducting disks with sizes
comparable or smaller than  coherence length or penetration depth
is quite different to that observed in macroscopic samples with
weak \cite{Fasano2005,Petrovic2009} or strong
\cite{Menghini2003,Demirdis} pinning. For instance, in disks
thinner than coherence length the confinement effect overwhelms
the inter-vortex interaction and vortices penetrate  in
ring-shaped shells  mimicking the sample edge
\cite{Schweigert98b}. However, inter-vortex interaction strongly
depends on the anisotropy of the  material and the resulting
mesoscopic vortex configuration can significantly change with the
degree of layerness of the material.

The case of mesoscopic vortex matter nucleated in the
extremely-layered high-$T_{\rm c}$ Bi$_{2}$Sr$_{2}$CaCu$_{2}$O$_{8
+ \delta}$ studied here is rather interesting since the system
presents a rich phase diagram. This is  due to the dominant effect
of thermal fluctuations and extreme anisotropy in samples with
weak disorder. The phase diagram of macroscopic as well as
mesoscopic \cite{Konczykowski2012} Bi$_{2}$Sr$_{2}$CaCu$_{2}$O$_{8
+\delta}$ vortex matter is dominated by a first-order transition
\cite{Pastoriza94,Zeldov95} at $T_{\rm FOT}$ between a solid phase
at low temperatures and a liquid \cite{Nelson1988} or decoupled
gas \cite{Glazman1991,Pastoriza1995} of pancake vortices at high
temperatures. Irrespective of the sample size, vortex penetration
in Bi$_{2}$Sr$_{2}$CaCu$_{2}$O$_{8 +\delta}$ is dominated by
Bean-Livingston surface barriers at high temperatures and bulk
pinning effects at low temperatures \cite{Morozov97,Wang02}. The
temperature  and measuring-time ranges in which every effect
dominates is determined by the thickness-to-width ratio of the
samples \cite{Morozov97,Wang02,Chikumoto1992}. This results in the
solid vortex phase presenting irreversible magnetic behavior
whereas the high-temperature  phase is magnetically reversible.
The transition between both phases is detected as a jump in the DC
local induction \cite{Pastoriza94} and as a frequency-independent
peak in the AC transmittivity \cite{Morozov1996,Konczykowski2006}.

In this work we study mesoscopic vortex matter penetrating into
micron-sized Bi$_{2}$Sr$_{2}$CaCu$_{2}$O$_{8 +\delta}$ disks by
means of local DC and AC micro-Hall-magnetometry measurements with
low-noise level. We detect features in the AC transmittivity and
third-harmonic signal at certain fields and ascribe them to the
abrupt entrance of single vortex-shells  on increasing field.

\begin{figure}[ttt]
\hspace{1.5cm}\includegraphics[width=0.8 \textwidth]{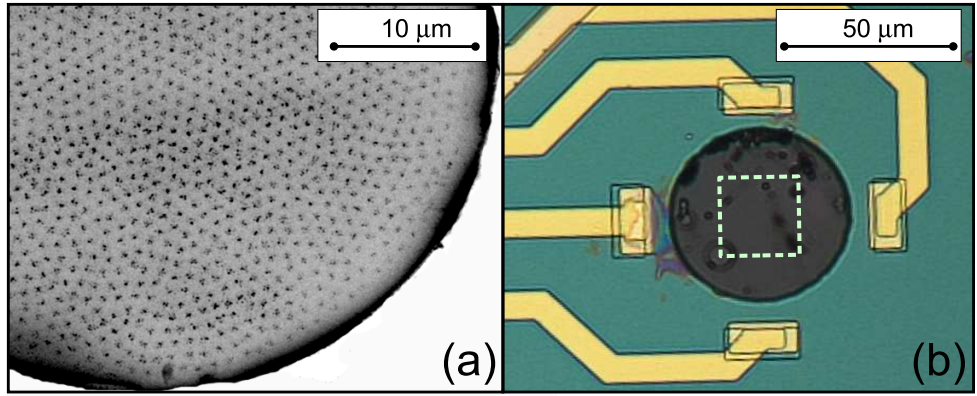}
\caption{\label{figure1} (a) Magnetic decoration image
\cite{Fasano2003} of the vortex matter nucleated in a
field-cooling process at 20\,Oe in a 50\,$\mu$m diameter
Bi$_{2}$Sr$_{2}$CaCu$_{2}$O$_{8 +\delta}$ disk.  (b) Typical disk
located on top of the $16 \times 16$\,$\mu$m$^{2}$ Hall sensor
used to measure (depicted with a dotted-white frame in real
size).}
\end{figure}

\section{Experimental}

We studied disks fabricated from macroscopic optimally-doped
Bi$_{2}$Sr$_{2}$CaCu$_{2}$O$_{8 +\delta}$ ($T_{\rm c}=90$\,K)
single-crystals grown by means of the traveling-solvent floating
zone technique \cite{Li94a}. Micron-sized disks of roughly
1\,$\mu$m thickness and diameter $d=50\,\mu$m as the one shown in
Fig.\,\ref{figure1} (a) were obtained by combining optical
lithography and physical ion-milling techniques \cite{Dolz10a}.
The disks were mounted with micron-precision manipulators onto
$2D$-electron-gas  Hall-sensors with active areas of $16 \times
16$\,$\mu$m$^{2}$ and glued with Apiezon N grease, see
Fig.\,\ref{figure1} (b).

The magnetic properties at individual-vortex scale nucleated in
field-cooling  were studied by means of magnetic decoration
imaging \cite{Fasano1999}. Local magnetization was measured
applying DC, $H$, and AC , $H_{\rm ac}$, fields parallel to the
$c$-axis. We simultaneously measure the first and third harmonics
of the AC induction by means of a digital-signal-processing
lock-in technique \cite{Konczykowski2012}. We obtained the
transmittivity, $T'$,  by normalizing the in-phase component of
the first harmonic signal \cite{Gilchrist1993}. This is a
magnitude extremely sensitive to discontinuities in the local
induction. The modulus of the third harmonic signal is similarly
converted to the magnitude $\mid T_{h3} \mid$
 that has a non-negligible value for
non-linear magnetic response at $H< H_{\rm irr}$, the
irreversibility field
 \cite{Gilchrist1993}.

\begin{figure}[ttt]
\hspace{0.8cm}\includegraphics[width=0.9\textwidth]{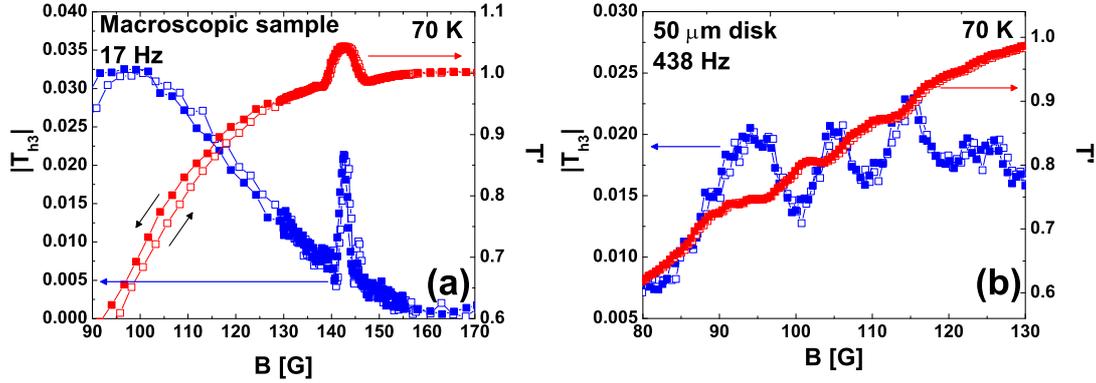}
\caption{\label{figure2} Field-evolution of the transmittivity
$T'$ (red points) and third-harmonic
 signal $\mid T_{h3} \mid$ (blue points) at 70\,K. Results for (a) the macroscopic sample from which were fabricated the (b)
 50\,$\mu$m diameter disks of  Bi$_{2}$Sr$_{2}$CaCu$_{2}$O$_{8 + \delta}$. The ripple $H_{\rm ac}$
fields have $\sim 1$\,Oe rms in magnitude and frequencies as
indicated. Peaks in $\mid T_{h3} \mid $ echo the oscillations in
$T'$. The open (full) points correspond to the warming (cooling)
branch as indicated by the black arrows.}
\end{figure}

\section{Results and discussion}

Figure \ref{figure1} (a) shows a zoom-in of a magnetic decoration
image revealing the  vortex arrangement in the disks for a
field-cooling experiment at 20\,Oe. At the edge of the sample
vortices arrange in a circular-shaped shell mimicking the disk
geometry. This structure is the result of the
nucleation of the vortex ensemble  in the high-temperature vortex
liquid and the subsequent freezing of the structure at the scale
of lattice spacing at the temperature at which bulk pinning sets
in \cite{Fasano2005}, namely $T_{\rm irr}(20\,Oe) = 85.3$\,K$\sim
T_{\rm FOT}$ for the 50\,$\mu$m disks
\cite{Konczykowski2012}.

For the flux-penetration  at fixed temperature experiments we
discuss from now on, vortices enter into the sample on increasing
field above the first-penetration field, $B_{\rm p}(T)$. In this
experiment, a snapshot of flux-penetration  would reveal a vortex
arrangement certainly different to the one shown in
Fig.\,\ref{figure1} (a), with an induction gradient towards the
sample center. However, Fig.\,\ref{figure1} (a) suggest that in
flux-penetration experiments vortices will enter into the sample
in circular-shaped shells. The entrance of these shells  can be
continuous or discretized, namely vortices gradually enter one by
one, or a circle of vortices suddenly jump into the sample.

In macroscopic Bi$_{2}$Sr$_{2}$CaCu$_{2}$O$_{8 + \delta}$ samples,
the first-order phase transition is detected as a sharp  peak in
AC transmittivity \cite{Morozov1996,Konczykowski2006}, a magnitude
sensitive to discontinuities in the magnetic induction
(proportional to the first derivative of the signal). Figure
\ref{figure2} (a) shows that for the macroscopic sample from which
were fabricated the disks this peak develops at $H_{\rm FOT} =
143$\,G at 70\,K. Peaks or oscillations in $T'$ are echoed as
peaks in $\mid T_{h3} \mid$, see Fig.\,\ref{figure2} (a), a
magnitude proportional to the third derivative of the signal. On
increasing field above $H_{\rm irr} \sim 160$\,G (for 17\,Hz),
$\mid T_{h3} \mid$ is below the experimental resolution indicating
a linear vortex magnetic response.

When vortices penetrate micron-sized disks the first-order phase
transition do persist at the same $H_{\rm FOT}$ than in
macroscopic samples \cite{Konczykowski2012}. In addition, if the
noise level and field-resolution of the measurements are improved,
additional features are detected in the $T'$ and $\mid T_{h3}
\mid$ curves for fields smaller than $H_{\rm FOT}$. On increasing
field at 70\,K vortices penetrate and oscillations in $T'$ are
observed in the range of 80 to 130\,G, see Fig.\,\ref{figure2}
(b). This contrasts with the flat $T'$ curve measured in this
field-range for the macroscopic Bi$_{2}$Sr$_{2}$CaCu$_{2}$O$_{8 +
\delta}$ sample. These features observed in the
micron-sized disks are due to discontinuities in the local
induction when vortices are entering (full points) or exiting
(open points) from the disks. The amplitude of the
oscillations in $T'$ for the disks are two-to-four times smaller
than the peaks measured in the macroscopic sample at $H_{\rm
FOT}$.

\begin{figure}[ttt]
\hspace{0.8cm} \includegraphics[width=0.9\textwidth]{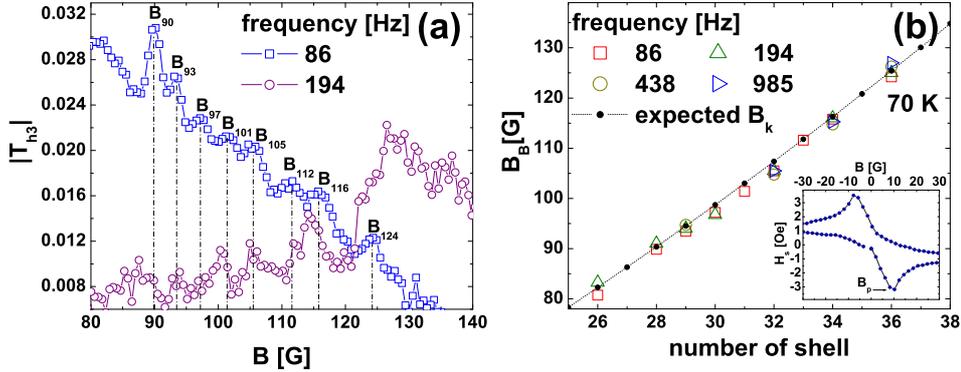}
\caption{\label{figure3} Third-harmonics signal of vortex matter
penetrating 50\,$\mu$m Bi$_{2}$Sr$_{2}$CaCu$_{2}$O$_{8 + \delta}$
disks at 70\,K. (a) Field-evolution of $\mid T_{h3} \mid$ for
$h_{ac}$ with different frequencies and a magnitude of 1.5\,Oe rms
(warming branches). The fields at which the peaks develop are
indicated as $B_{\rm B}$. (b) Field-location of $\mid T_{h3} \mid$
peaks (open color points) and of the expected single-shell
discretized penetration fields (full black points and dotted
line), as a function of the single-shell number.  Insert: Local DC
magnetization loop at 70\,K with the first-penetration field
indicated.}
\end{figure}

Figure \ref{figure2} (b) shows that the oscillations
in $T'$ are echoed as peaks in $\mid T_{h3} \mid$. We were able to
detect these peaks in the temperature-range between 70 to 90\,K.
 For the 50\,$\mu$m disks at 70\,K, the $\mid T_{h3}
\mid$ peaks are detected  between 80 and 130\,G, for a wide range
of frequencies, see the examples of Fig.\,\ref{figure3} (a).
We identify the particular fields at which the peaks
develop with $B_{\rm B}$, with $B$ the field-location of the
maximum. For instance, Fig.\,\ref{figure3}(a) depicts peaks
$B_{90}$ to $B_{124}$. The
fields at which $\mid T_{h3} \mid$ becomes negligible on
increasing vortex density rises with frequency, namely $H_{\rm
irr}$ moves upward with frequency \cite{Konczykowski2012}.

On sweeping up field, flux-penetration in the 50\,$\mu$m disks
starts at  $B_{\rm p} = 10$\,G with the first single-shell of
vortices nucleating at the sample edge. The $B_{\rm p}$ field is
obtained from DC local magnetization loops as indicated in the
insert to Fig.\,\ref{figure3} (b). Further increasing the field
might entail either a continuous entrance of vortices, or a
discretized process in which vortices do not enter until the
increment in $H$ is such that its vorticity allows the formation
of a new single-shell of vortices. In the latter case, the new
$k$-th single-shell of vortices jumps  into the sample for a given
$B_{\rm k}$ field. We propose that the  peaks observed in $\mid
T_{h3}\mid$ concomitant with the oscillations in $T'$ are the
fingerprint of this discretized entrance of vortices on increasing
$H$, namely that $B_{\rm B} \sim B_{\rm k}$ for given $k$ and $\rm
B$. If vortices would penetrate the sample individually in a
continuous way, no such sharp features in $T'$ and $\mid
T_{h3}\mid$ are expected.

We can also consider a very simple geometrical analysis in order
to relate the observed $B_{\rm B}$ features with the entrance of a
$k$-th shell. By considering the measured $B_{\rm p}$, we estimate
the number of vortices suddenly penetrating in the first
single-shell ($k=1$) as $n_{1} = \pi B_{\rm p} d^{2}/4\Phi_{0}$,
with $\Phi_{0}$ the flux quantum. We then calculate the required
increment in field in order to produce the jumping-in of the
second single-shell ($k=2$), $\Delta B_{2}$. In order to do so we
consider that $n_{2} = (B_{\rm p} + \Delta B_{2})(\pi
d^{2})/4\Phi_{0}$ and $n_{2} = n_{1} + \Delta n$ with the number
of vortices in the second single-shell $\Delta n = (\pi d)/a_{2}$
with $a_{2}$ the vortex spacing. We proceed in the same fashion
for the following single-shells, and obtained the $B_{k}$ curve
shown in Fig.\,\ref{figure3} (b) as a function of the
single-shells ordinal number $k$. The vortex spacing is considered
in a very rough approximation as
$a_{k}=1.075\sqrt{\Phi_{0}/B_{k-1}}$ with the special case of
$a_{2}=1.075\sqrt{\Phi_{0}/B_{p}}$. This is a very simple analysis
but since we do not have direct imaging evidence on the
penetration of vortices within a single-shell, we can not suggest
a more sophisticated model.

We then identified all the observed $B_{\rm B}$ peaks from $\mid
T_{h3}\mid$ measurements at several frequencies in the range of 80
to 130\,G. For every $B_{\rm B}$ we then find the closest
magnetic-field value for which is expected the entrance of a
single-shell of vortices, $B_{\rm k}$, and construct the curves
shown in Fig.\,\ref{figure3} (b). Namely, we assume that the peaks
in $\mid T_{h3}\mid$ are located at the particular field-values
$B_{\rm k}$ such that its vorticity allows a single-shell of
vortices with ordinal number $k$ penetrating into the sample. In
spite of the simplicity of this analysis, the agreement between
the experimental points and the values expected from this
geometrical model is quite remarkable.

This interpretation could be challenged by arguing that the
oscillations in $T'$ come from inhomogeneities in the sample
giving different local values of $H_{\rm FOT}$ at a fixed
measurement temperature. This seems not to be the case since these
features in $T'$ have a different shape than the paramagnetic
peaks associated to the transition at  $H_{\rm FOT}$, and they are
observed in a different field-range than this transition.
Nevertheless, we did an extra experimental check in order to
verify our interpretation. In the case of the  $T'$ and
concomitant $\mid T_{h3}\mid$ peaks associated to the first-order
transition in macroscopic samples, applying an in-plane field
produces a  shifting of the peaks towards smaller fields
\cite{Konczykowski2006}. When we applied such a field in our
50\,$\mu$m disks we found that the field-location of the $B_{\rm
B}$  peaks in $\mid T_{h3}\mid$ does not change appreciably.
Therefore with these experiments we strengthen our interpretation
of the peaks in $\mid T_{h3}\mid$ being the fingerprint of the
discretized entrance of single-shells of vortices. Vortex imaging experiments on increasing field in films
with strong periodic pinning potentials show that the entrance of
a vortex shell is not performed abruptly but in steps following
the terrace critical state \cite{Silhanek2011}. However, in our
case the samples do not have a periodic pinning potential and such
a penetration state is not expected. Nevertheless, whether  the
penetration is performed with all vortices of a single-shell
entering suddenly at the same field, or slightly gradually in the
small range of fields the $\mid T_{h3}\mid$ peaks widen, can not
be ascertained from our experiments. 

Finally, we find that on further increasing the field above
130\,Oe the oscillations in $T'$ as well as the peaks in the
third-harmonic signal progressively faint. This suggest that in
this field-range the
 field-steps required to allow the discretized entrance of
single-shell of vortices become smaller than the field-resolution
of our experiments of roughly 0.1\,G.  Therefore this
phenomenology is resolution-limited and we can not ascertain
whether there is a discrete-to-continuum crossover of flux
penetration in micron-sized vortex matter.

\section{Conclusions}

Low-noise level and high-resolution  local magnetic measurements,
as well as single-vortex imaging, allowed us to detect the
entrance of quite-likely ring-shaped single-shells of vortices
penetrating micron-sized samples. Therefore for the mesoscopic
vortex matter nucleated in the extremely-layered
Bi$_{2}$Sr$_{2}$CaCu$_{2}$O$_{8 +\delta}$ superconductor,
flux-penetration at low fields follows a discretized process in
which the confinement plays a relevant role on imposing the
geometry of the sample for vortex entrance. This effect is no
longer detected for a system with more  than 13.000 vortices
 indicating inter-vortex interaction becomes more
relevant.


\section*{References}
\begin{thebibliography}{9}




\bibitem{Moshchalkov95} V. V. Moshchalkov, L. Gielen, C. Strunk, R. Jonckheere, X. Qiu
\textit{et al.}, Nature \textbf{373}, 319 (1995).


\bibitem{Geim97} A. K. Geim, I. V. Grigorieva, S. V. Dubonos, J. G. S. Lok, J. C.
Maan \textit{et al.}, Nature \textbf{390}, 256 (1997).

\bibitem{Schweigert98b} V. A. Schweigert, F. M. Peeters, and P. S. Deo, Phys. Rev. Lett.
\textbf{81}, 2783 (1998).

\bibitem{Palacios98} J. J. Palacios, Phys. Rev. B \textbf{58}, R5948 (1998).

\bibitem{Bruyndoncx99} V. Bruyndoncx, L. Van Look, M. Verschuere, and V. V.
Moshchalkov, Phys. Rev. B \textbf{60}, 10468 (1999).

\bibitem{Cabral04} L. R. E. Cabral, B. J. Baelus, and F. M. Peeters, Phys. Rev. B
\textbf{70}, 144523 (2004).

\bibitem{Fasano2005} Y. Fasano, M. De Seta, M. Menghini, H.
Pastoriza and F. de la Cruz, Proc. Nat. Acad. Sci. \textbf{102},
3898 (2005).

\bibitem{Petrovic2009} A. P. Petrovic, Y. Fasano, R. Lortz, C. Senatore, A. Demuer
\textit{et al.}, Phys. Rev. Lett. \textbf{103}, 257001 (2009).

\bibitem{Menghini2003} M. Menghini, Y. Fasano, F. de la Cruz, S. S. Banerjee \textit{et al.}, Phys. Rev. Lett. \textbf{90}, 147001
(2003).

\bibitem{Demirdis} S. Demirdis, C.J. van der Beek, Y. Fasano, N. R. Cejas Bolecek \textit{et al.},
Phys. Rev. B \textbf{84}, 094517 (2011).

\bibitem{Konczykowski2012} M. Konczykowski, Y. Fasano, M. I. Dolz, H. Pastoriza, V. Mosser, and M.
Li,  arXiv:1212.4564 (2012).

\bibitem{Pastoriza94} H. Pastoriza, M. F. Goffman, A. Arribere, and F. de la Cruz,
Phys. Rev. Lett. \textbf{72}, 2951 (1994).

\bibitem{Zeldov95} E. Zeldov, D. Majer, M. Konczykowski, V. B. Geshkenbein, V. M.
Vinokur \textit{et al.}, Nature \textbf{375}, 373 (1995).

\bibitem{Nelson1988} D. R. Nelson, Phys. Rev. Lett. \textbf{60}, 1973 (1988).

\bibitem{Glazman1991} L. I. Glazman, and A. E. Koshelev, Phys. Rev. B \textbf{43}, 2835 (1991).

\bibitem{Pastoriza1995} H. Pastoriza and P. H. Kes, Phys. Rev. Lett. \textbf{75}, 3525 (1995).

\bibitem{Morozov97} N. Morozov, E. Zeldov, M. Konczykowski, and R.A. Doyle, Physica C
\textbf{291}, 113 (1997).

\bibitem{Wang02} Y. M. Wang, A. Zettl, S. Ooi and T. Tamegai Phys. Rev. B, \textbf{65},
184506(2002).

\bibitem{Chikumoto1992} N. Chikumoto, M. Konczykowski, N. Motohira, and A. P. Malozemoff,
Phys. Rev . Lett. \textbf{69} 1260 (1992).

\bibitem{Morozov1996} N. Morozov, E. Zeldov, D. Majer and M. Konczykowski, Phys. Rev. B \textbf{54}, R3784 (1996).

\bibitem{Konczykowski2006} M. Konczykowski, C. J. van der Beek, A. E. Koshelev, V. Mosser \textit{et al.}, Phys. Rev. Lett.
\textbf{97}, 237005 (2006).

\bibitem{Li94a} T. W. Li, P.H. Kes, N.T. Hien, J. J. M. Franse and A.
A. Menovsky, J. Cryst. Growth \textbf{135}, 481 (1994).

\bibitem{Dolz10a} M. I. Dolz, A. B. Kolton and H. Pastoriza, Phy. Rev. B \textbf{81}, 092502 (2010).

\bibitem{Fasano2003} Y. Fasano, M. De Seta, M. Menghini, H. Pastoriza, and F. De la Cruz,  Solid State
Commun. \textbf{128}, 51 (2003).

\bibitem{Fasano1999} Y. Fasano, J. Herbsommer, and F. de la Cruz, Phys. Stat. Sol. (b)
\textbf{215}, 563 (1999).

\bibitem{Gilchrist1993} J. Gilchrist, and M. Konczykowski, Phys. C \textbf{212}, 43
(1993).

\bibitem{Silhanek2011} A. V. Silhanek \textit{et al.} Phys. Rev. B \textbf{83}, 024509 (2011).



\end{thebibliography}
\end{document}